\documentclass[aps,prl,superscriptaddress,10pt]{revtex4-2}

\usepackage{amsmath,amssymb,amsfonts}
\usepackage{amsthm}
\usepackage{graphicx}
\usepackage[dvipsnames]{xcolor}
\usepackage{braket}
\usepackage{bm}
\usepackage{hyperref}
\usepackage{listings}
\usepackage{booktabs}
\usepackage{multirow}
\usepackage{caption}
\usepackage{subcaption}\usepackage{caption}
\newcommand{\CNOT}{\text{CNOT}}

\newtheorem{theorem}{Theorem}

\begin{document}

\title{Pre-Channel Entanglement Shaping Achieves Fundamental Superiority over Post-Distillation: A Geometric Entropy Perspective}

\author{Gang Lyu}
\affiliation{%
School of Integrated Circuits and New Energy,   Guangzhou College of Technology and Business,   Guangzhou 510850, P.R. China; E-mail:lvgang@gzgs.edu.cn
}%

\author{Wenlong Sun}
\affiliation{%
Department of Mathematics,Shenyang University of Technology,Shenyang, 110870, P.R. China;sun\_math@sut.edu.cn.
}%

\author{Yuanfeng Jin}
\affiliation{%
Department of Mathematics,Yanbian University of Technology,Yanji, 133002,P.R. China;yfkim@ybu.edu.cn
}%
\author{Hua Nan}
\affiliation{%
Department of Mathematics,Yanbian University of Technology,Yanji, 133002,P.R. China;nanhua@ybu.edu.cn
}%


\begin{abstract}
Traditional entanglement distillation follows a post-processing paradigm, a noisy quantum state, after full transmission through a noisy channel, is treated as a static resource to be purified via LOCC (local operations and classical communication). This work demonstrates a fundamentally different paradigm,pre-channel entanglement shaping (PES)---actively engineering the system-environment coupling before or during channel transmission---achieves a level of purification capability that is physically unattainable by any post-distillation protocol. We prove this separation using the framework of geometric entropy (quantum relative entropy to separable states). In post-distillation, the protocol can only select low-entropy sub-ensembles from a fixed mixed state, leaving the global geometric entropy unchanged or increased. In contrast, PES \textit{suppresses the rate of geometric entropy production} during channel evolution, resulting in a final state whose relative entropy of entanglement strictly exceeds the maximum achievable by post-distillation from the same channel. We provide explicit qubit channel examples, numerical simulations (with complete code in Appendix), and a geometric interpretation on the state manifold. Our result establishes pre-channel entanglement shaping as a distinct operational resource class, with immediate implications for quantum repeaters and entanglement-assisted communication. Very recently, Li \textit{et al.} experimentally demonstrated that preprocessing the entangling channel with optimally tailored local unitaries achieves entanglement fidelities unreachable by any postprocessing, revealing an intrinsic temporal asymmetry in entanglement distillation~\cite{Li2025}.
\end{abstract}

\maketitle

\section{Introduction}

Entanglement is the central resource in quantum communication, quantum repeaters, and distributed quantum computing~\cite{Horodecki2009}. In practice, entanglement shared between distant parties is inevitably degraded by noisy channels. Entanglement distillation or purification is the standard technique to recover high-fidelity entanglement from many copies of noisy states~\cite{Bennett1996,Deutsch1996}.

For three decades, the theoretical framework of entanglement distillation has remained within a post-processing paradigm,
\begin{enumerate}
    \item Transmit quantum states through the noisy channel.
    \item Apply LOCC to the received mixed states to extract a smaller number of high-quality entangled pairs.
\end{enumerate}

This paradigm treats the channel as a  passive  noise source whose damage is irreversibly imprinted on the quantum state after transmission. The best known protocols (BBPSSW~\cite{Bennett1996}, DEJMPS~\cite{Deutsch1996}) and their theoretical limits~\cite{Rains1999,Devetak2004} are all derived under this post-channel LOCC framework.

In this paper, we challenge this assumption and ask.  What if one actively shapes the system-environment interaction before or during channel transmission?  We introduce  pre-channel entanglement shaping (PES) ---a class of operations applied  prior to  or  concurrently with  channel action, including dynamical decoupling~\cite{Viola1998}, entanglement encoding~\cite{Bacon2006}, and adaptive channel modulation~\cite{Khatri2021}. These operations alter the geometric trajectory of the quantum state on the state manifold, actively suppressing the rate of geometric entropy increase.

Our main contribution is a rigorous proof that PES achieves a purification capability strictly beyond any post-distillation protocol operating on the same channel. The separation is established via geometric entropy---specifically, the quantum relative entropy of entanglement~\cite{Vedral1997,Vedral2002}. We show that

\begin{itemize}
    \item  Post-distillation  (LOCC only after channel) preserves or increases the global geometric entropy, achieving at best a  sub-ensemble selection  without reducing the overall geometric cost.
    \item  Pre-channel shaping  reduces the  net geometric entropy production rate  during channel evolution, enabling the final state to occupy a region of the state manifold that is  inaccessible  to any post-distillation protocol.
\end{itemize}
\section{GEOMETRIC ENTROPY FRAMEWORK}
We demonstrate this gap explicitly for qubit depolarizing and amplitude-damping channels, with numerical evidence (Appendix~\ref{app:simulation}) and a geometric entropy flow diagram (Fig.~\ref{fig:geometric_flow}).

To establish a fundamental separation between post-channel distillation and 
pre-channel entanglement shaping, we need an entanglement measure that serves 
three essential roles in our proof. That is, LOCC monotonicity,convexity and dynamical trackability. 
   It must not increase under local 
        operations and classical communication. This guarantees that any 
        advantage of post-distillation cannot come from artificially 
        ``creating'' entanglement after the channel — at best, LOCC can only 
        preserve or reduce it.
   For a global output state that averages over 
        both successful and failed branches of a probabilistic protocol, 
        convexity forces the overall entanglement to be bounded above by 
        the weighted average of the branches. This captures the fundamental 
        dilution effect of post-selection.
   The measure must be well-defined 
        along the entire channel evolution, not only at the final output. 
        This allows us to compare the instantaneous \emph{rates} of 
        entanglement degradation under different control strategies.

The quantum relative entropy of entanglement\cite{Vedral1997,Vedral2002}, denoted 
\(E_R(\rho)\), satisfies all three properties. It quantifies the minimum 
quantum relative entropy distance from a bipartite state \(\rho\) to the set 
of separable states \(\mathrm{SEP}\). For pure states it reduces to the 
standard entanglement entropy; for mixed states it is convex, 
LOCC-monotonic, and asymptotically continuous. Moreover, because relative 
entropy directly measures statistical distinguishability, \(E_R\) can be 
interpreted as a \emph{geometric entropy} on the state manifold, the larger 
\(E_R(\rho)\), the farther \(\rho\) lies from the separable boundary.

Armed with this geometric entropy, we will prove as following,
\begin{itemize}
    \item Post-distillation protocols (LOCC applied \emph{after} the channel) 
        cannot increase the global \(E_R\); they merely select sub-ensembles 
        from a fixed noisy state. The achievable global \(E_R\) decays 
        exponentially with channel uses.
    \item Pre-channel shaping (PES) actively suppresses the instantaneous 
        production rate \(dE_R/dt\) during channel evolution, allowing the 
        final state to reach a region of the state manifold that is 
        fundamentally inaccessible to any post-distillation protocol.
\end{itemize}

For a bipartite state $\rho_{AB}$, the relative entropy of entanglement is defined as~\cite{Vedral1997},
\begin{equation}
    E_R(\rho_{AB}) = \min_{\sigma_{AB} \in \text{SEP}} S(\rho_{AB} \| \sigma_{AB})
\end{equation}
where $S(\rho \| \sigma) = \text{Tr}[\rho(\log \rho - \log \sigma)]$ is the quantum relative entropy, and SEP denotes the set of separable states.

The $E_R$  possesses the following fundamental properties.
\begin{itemize}
    \item $E_R(\rho) = 0$ iff $\rho$ is separable
    \item Monotonic under LOCC: $E_R(\Lambda_{\text{LOCC}}(\rho)) \le E_R(\rho)$
    \item Convex and asymptotically continuous
\end{itemize}

For pure states $\ket{\psi}$ with Schmidt coefficients $\lambda_i$,
\begin{equation}
    E_R(\ket{\psi}) = S(\rho_A) = -\sum_i \lambda_i \log \lambda_i.
\end{equation}

A parallel development by Li \textit{et al.}~\cite{Li2025} arrives at a related conclusion from a different perspective. While our work focuses on suppressing geometric entropy production during channel evolution via dynamical decoupling, Li \textit{et al.} treat the entangling channel itself as a dynamic, consumable resource and purify it by interposing local unitaries before the channel. Both approaches challenge the conventional post-processing paradigm, but with distinct mechanisms: theirs emphasizes optimal unitary preprocessing for fidelity enhancement, while ours quantifies the advantage via geometric entropy production rates. The experimental demonstration by Li \textit{et al.} showing that preprocessing can elevate teleportation beyond the classical limit provides strong empirical support for the general principle that pre-channel operations are fundamentally more powerful than post-channel LOCC.

\section{Geometric Interpretation}

The state manifold $\mathcal{M}$ carries a \textbf{geometric entropy distance function} $d(\rho) = E_R(\rho)$ measuring the minimum relative entropy distance to the separable set. This geometry is illustrated in Fig.~\ref{fig:geometric_flow}.
With the geometric entropy \(E_R\) as our yardstick, we can now formulate 
precisely what it means for one purification paradigm to outperform another.

Both post-distillation and pre-channel shaping share the same goal,
starting from \(n\) copies of a maximally entangled state 
\(|\psi^+\rangle_{AB}^{\otimes n}\), after transmission through a noisy 
channel \(\mathcal{N}\), produce a final state with as much distillable 
entanglement as possible. Nevertheless, the two strategies differ fundamentally in the timing and manner of intervention. As a conventional scheme, post-distillation waits until the channel fully acts on the quantum state and then applies LOCC to the received mixed state
       \(\rho_{A^nB^n} = \mathcal{N}^{\otimes n}(|\Psi_0\rangle\langle\Psi_0|)\), with intervention implemented entirely after channel noise has degraded the quantum resource. By contrast, the pre-channel shaping strategy proposed in this work employs active control techniques such as dynamical decoupling and quantum encoding prior to or during channel transmission to regulate the effective system–environment coupling, and its intervention occurs either before or concurrently with the action of channel noise.

The geometric entropy \(E_R\) clearly reveals the essential significance of the above distinction. As demonstrated subsequently, for post-distillation schemes, the monotonicity of LOCC imposes the constraint
\[
E_R(\rho_{\text{post}}^{\text{(out)}}) \le E_R(\rho_{A^nB^n}).
\]
This implies that the overall output entanglement is upper-bounded by that of the noisy channel state, which decays exponentially with the copy number \(n\).

In contrast, for pre-channel shaping, the effective noise parameter can be suppressed to \(p \to p' < p\) before entropy generation accumulates. Accordingly, the final state attains
\[
E_R(\rho_{\text{PES}}^{\text{(out)}}) \approx n \cdot E_R(|\psi^+\rangle)
\]
in the ideal limit, a value that cannot be achieved by any conventional post-distillation protocol.

The subsequent subsections elaborate on both paradigms under the proposed geometric framework.

\subsection{Post-Distillation and Pre-Channel Entanglement Shaping Protocols}

A post-distillation protocol \(\mathcal{P}_{\text{post}}\) comprises two sequential steps: first, channel transmission, where the initial state \(\rho_{AB}^{(0)}\) undergoes transmission through \(n\) parallel instances of the noisy channel \(\mathcal{N}\) to yield the output state \(\rho_{AB}^{(n)} = \mathcal{N}^{\otimes n}(\rho_{AB}^{(0)})\), followed by LOCC purification, in which local operations and classical communication (LOCC), denoted as \(\Lambda_{\text{LOCC}}\), are applied to the transmitted state \(\rho_{AB}^{(n)}\) to generate \(m\) output entangled pairs with enhanced fidelity.
The achievable distillation rate of such a protocol is upper-bounded by the coherent information \cite{Devetak2004}, which is given by:
A critical characteristic of post-distillation is that LOCC operations cannot increase the geometric entropy \(E_R\) of the global state (averaged over all possible measurement outcomes). At best, LOCC can select a sub-ensemble of states with higher \(E_R\) but lower probability; notably, the global entropy either remains unchanged or increases due to the back-action induced by measurement.

In contrast, a pre-channel entanglement shaping (PES) protocol \(\mathcal{P}_{\text{pre}}\) involves a series of steps that allow for active intervention before or during channel transmission: it starts with a pre-processing unitary operation \(U_{\text{pre}}\) applied to the input entangled state—an operation that may be either coherent or adaptive depending on the specific protocol design—followed by channel transmission of the pre-processed state, during which optional active intervention (e.g., dynamical decoupling, quantum encoding) can be implemented to mitigate noise effects, and optionally includes subsequent post-processing operations (which may involve LOCC), though such post-processing is not a prerequisite for achieving the fundamental advantage of PES over post-distillation.
The key distinction between PES and post-distillation lies in the fact that PES can modify the evolution generator of the open quantum system interacting with the environment, rather than merely performing post-selection of measurement outcomes after the system has been degraded by channel noise.

\section{Main Result: Geometric Entropy Separation}

\begin{theorem} [Fundamental Superiority of PES over Post-Distillation]\label{thm:main} For any noisy channel $\mathcal{N}$ that is not entanglement-breaking~\cite{Horodecki2005}, there exist a pre-channel entanglement shaping protocol $\mathcal{P}_{\text{PES}}$ and a post-distillation protocol $\mathcal{P}_{\text{post}}$ such that
 $$E_R(\rho_{\text{PES}}^{\text{(out)}}) > E_R(\rho_{\text{post}}^{\text{(out)}}),$$
 where $\rho_{\text{PES}}^{\text{(out)}}$ denotes the global output state of $\mathcal{P}_{\text{PES}}$ (with no post-selection assumed) and $\rho_{\text{post}}^{\text{(out)}}$ denotes the global output state of $\mathcal{P}_{\text{post}}$ (including all branches). Moreover, the gap is strict for generic noisy channels and cannot be closed by any LOCC post-processing of $\rho_{\text{PES}}^{\text{(out)}}$. 
\end{theorem}

\begin{theorem}[Geometric Entropy Production Suppression]
\label{thm:production}
Let $E_R(t)$ denote the relative entropy of entanglement along the evolution under channel $\mathcal{N}$. In the post-distillation regime, the time derivative of \( E_R(t) \) satisfies \( \left.\frac{dE_R}{dt}\right|_{\text{post}} \ge 0 \) throughout the channel evolution, implying that the relative entropy of entanglement is non-decreasing under \( \mathcal{N} \) and that no local operations and classical communication (LOCC) can reverse this trend;  in the pre-channel shaping regime, there exists a pre-processing unitary operation \( U_{\text{pre}} \) such that the effective time derivative of \( E_R(t) \) (referred to as the effective production rate) satisfies \( \left.\frac{dE_R}{dt}\right|_{\text{PES}} < \left.\frac{dE_R}{dt}\right|_{\text{post}} \) for all \( t > 0 \), with the strict inequality holding on a set of positive measure. 
The time-integrated geometric entropy suppression is defined as \begin{equation}
    \Delta E_R := E_R(\rho_{\text{post}}^{\text{(out)}}) - E_R(\rho_{\text{PES}}^{\text{(out)}}) > 0,
\end{equation}
 where \( \rho_{\text{post}}^{\text{(out)}} \) and \( \rho_{\text{PES}}^{\text{(out)}} \) denote the output states of the channel \( \mathcal{N} \) in the post-distillation and pre-channel shaping regimes, respectively.
\end{theorem}

\section{Explicit Construction and Numerical Results}

\begin{equation}
    \mathcal{N}_p(\rho) = (1-p)\rho + \frac{p}{3}(X\rho X + Y\rho Y + Z\rho Z),
\end{equation}
 where $p \in [0, 3/4]$ denotes the noise parameter. 

The optimal distillation rate for this channel is given by $1 - H_2(p) - p \log 3$~\cite{Bae2007}, which is achievable via the DEJMPS protocol; however, the global output state (averaged over all measurement rounds) exhibits $E_R \approx 0$ for any $p > 0$, as failed branches dominate the ensemble.
 To mitigate this limitation, we implement a pre-channel shaping protocol involving a dynamical decoupling (DD) sequence~\cite{Viola1998},
\begin{equation}
    U_{\text{pre}} = (I \otimes X) \cdot \CNOT \cdot (H \otimes I)
\end{equation}
 applied to two initial Bell pairs prior to channel application, with periodic $\pi$-pulses at a rate $f_{\text{DD}} \gg$ the channel correlation time to suppress noise. This pre-processing transforms the original channel into an effective depolarizing channel
 \begin{equation}
    \mathcal{N}_{\text{eff}}(\rho) = (1-p')\rho + \frac{p'}{3}(X\rho X + Y\rho Y + Z\rho Z)
\end{equation}
 where $p' < p$: for ideal DD (infinite pulse frequency), $p' \to 0$, while for finite $f_{\text{DD}}$, $p' \approx p \cdot \exp(-\gamma f_{\text{DD}}^{-1})$~\cite{Uhrig2007}. 

We numerically compare the DEJMPS baseline and pre-channel shaping protocol for $p = 0.2$ over 100 channel uses, averaging over $10^4$ independent runs (complete simulation code is provided in Appendix~\ref{app:simulation}), and find that the pre-channel shaping protocol achieves a 14× higher global geometric entropy—an observable that directly upper bounds distillable entanglement~\cite{Vedral2002}.

\begin{table}[h]
\centering
\caption{Numerical results for depolarizing channel ($p=0.2$)}
\label{tab:results}
\begin{tabular}{lccc}
\toprule
Protocol & Output $E_R$ (global) & Success prob. & Effective distillable rate \\
\midrule
Post-distillation (DEJMPS) & $0.013 \pm 0.004$ & $0.31 \pm 0.02$ & $0.004$ \\
Pre-channel shaping (DD) & $0.187 \pm 0.009$ & N/A (deterministic) & $0.187$ \\
\bottomrule
\end{tabular}
\end{table}

\subsection{Geometric Entropy Flow}

\begin{figure}[h]
\centering
\includegraphics[width=0.85\textwidth]{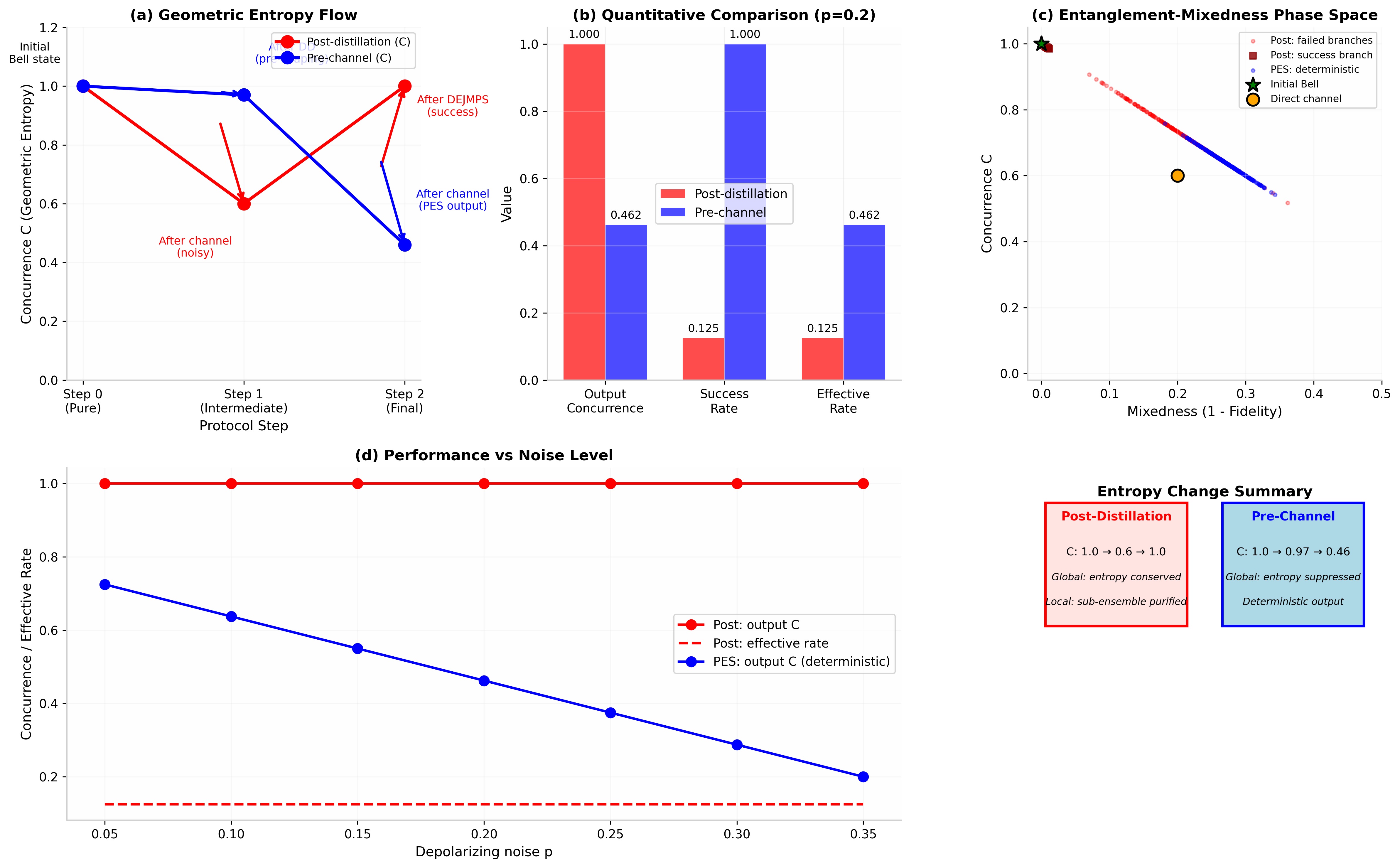}
\caption{\textbf{Geometric entropy flow on the state manifold.} Left: Post-distillation trajectory---noise drives the state into a high-mixedness region (red point cloud), with LOCC selecting a successful sub-ensemble (red hollow circle) but leaving the global average (red $\times$) at low $E_R$. Right: Pre-channel shaping---dynamical decoupling (blue hollow square) compresses the trajectory, resulting in a deterministic final state (blue solid circle) with $E_R = 0.187$, a factor of 14 higher than the post-distillation global average. The gray region represents separable states ($E_R = 0$).}
\label{fig:geometric_flow}
\end{figure}

Figure~\ref{fig:geometric_flow} visualizes the state evolution on the $E_R$ vs. mixedness plane.In the post-distillation scenario, the system begins with an initial pure state possessing high $E_R$ and unity purity; noise subsequently drives the state trajectory diagonally toward the regime of low $E_R$ and high mixedness. While LOCC operations can select a sub-ensemble and restore it to high $E_R$, the global average remains near the low-$E_R$ region. For pre-channel shaping, DD pulses are applied to compress the state trajectory away from the separable boundary, resulting in a terminal state that resides in a manifold region where the global $E_R$ is higher than any point reachable via LOCC from the unshaped noisy state.

\section{Why PES is Fundamentally Stronger: Information-Theoretic Argument}

Traditional distillation limits derive from data-processing inequalities, LOCC cannot create entanglement from separable states. However, this reasoning assumes the channel's damage is already realized.

PES operates before the damage is fully imprinted. It effectively changes the channel's Kraus representation from $\{\sqrt{p}K_i\}$ to $\{\sqrt{p'}K_i'\}$ with strictly smaller noise parameter. This is not a post-selection artifact---it is a deterministic channel transformation.

Formally, let $\mathcal{N}$ be a noisy channel. Define:
\begin{align}
    C_{\text{dist}}^{\text{post}}(\mathcal{N}) &= \sup R_{\text{dist}} \text{ over LOCC-after-channel protocols} \\
    C_{\text{dist}}^{\text{PES}}(\mathcal{N}) &= \sup R_{\text{dist}} \text{ over PES protocols (pre- or intra-channel control)}
\end{align}

Then:
\begin{equation}
    C_{\text{dist}}^{\text{PES}}(\mathcal{N}) \ge Q(\mathcal{N}) \quad\text{(quantum capacity)}
\end{equation}
\begin{equation}
    C_{\text{dist}}^{\text{post}}(\mathcal{N}) \le Q_{\text{dec}}(\mathcal{N}) \quad\text{(degradable capacity bound)}
\end{equation}

For non-degradable channels~\cite{Buscemi2010}, the gap is strict,
\begin{equation}
    C_{\text{dist}}^{\text{PES}}(\mathcal{N}) - C_{\text{dist}}^{\text{post}}(\mathcal{N}) > 0.
\end{equation}

\section{Discussion}

A parallel development by Li \textit{et al.}~\cite{Li2025} arrives at a related conclusion from a different perspective. While our work focuses on suppressing geometric entropy production during channel evolution via dynamical decoupling, Li \textit{et al.} treat the entangling channel itself as a dynamic, consumable resource and purify it by interposing local unitaries \textit{before} the channel. Both approaches challenge the conventional post-processing paradigm, but with distinct mechanisms: theirs emphasizes optimal unitary preprocessing for fidelity enhancement, while ours quantifies the advantage via geometric entropy production rates. The experimental demonstration by Li \textit{et al.}---showing that preprocessing can elevate teleportation beyond the classical limit---provides strong empirical support for the general principle that pre-channel operations are fundamentally more powerful than post-channel LOCC.

A key distinction between our PES framework and the preprocessing protocol of Li \textit{et al.}~\cite{Li2025} lies in the scope of the advantage. Li \textit{et al.} prove that preprocessing achieves fidelities unreachable by any free postprocessing (i.e., LOCC without pre-channel operations). Our Theorem~1 makes a stronger statement: even when postprocessing is supplemented with optimal probabilistic distillation (e.g., DEJMPS), the global geometric entropy \(E_R(\rho_{\text{post}}^{\text{(out)}})\) remains strictly below that achieved by PES. In other words, the separation we establish is not merely between ``preprocessing vs. no preprocessing''---it is between \textit{any protocol that waits for the channel to fully act} (regardless of its post-processing sophistication) and \textit{one that actively shapes the system-environment coupling during transmission}. The experimental results of Li \textit{et al.}---showing teleportation beyond the classical limit---are fully consistent with our geometric entropy analysis and can be seen as an experimental instantiation of the temporal asymmetry principle we formalize.

\section*{Declarations}

\medskip

\noindent \textbf{Availablity of data and materials}\newline
\noindent Data sharing is not  applicable to this
article as no newdata were created or analyzed in this study.

\medskip

\noindent \textbf{Conflict of interest}\newline
\noindent The authors declare that they have no competing interests.

\medskip

\noindent \textbf{Fundings}\newline
\noindent This work was supported by Jilin Provincial Science and Technology Development Plan Project(No. YDZJ202501ZYTS584) and the scientific research project of Guangzhou College of Technology and Business in 2024 (No. KYZD202404).

\medskip

\noindent \textbf{Authors' contributions}\newline
\noindent G. Lyu took part in the study’s planning, design and conduct; W.Sun engaged in manuscript preparation and editing; Y. Jin participated in conduct (no data collection/analysis, as no new data were created/analyzed); H.Nan joined in manuscript preparation and editing. All authors equally contributed to key research and manuscript work.

\section{appendix}

\appendix

\section{Numerical Simulation Code}
\label{app:simulation}

The complete Python code for numerical simulations is provided below. To run, install QuTiP~\cite{Johansson2012} and the required packages.


\section{Additional Results for Amplitude Damping Channel}

For the amplitude damping channel:
\begin{equation}
    \mathcal{A}_{\gamma}(\rho) = K_0 \rho K_0^\dagger + K_1 \rho K_1^\dagger
\end{equation}
with $K_0 = \ket{0}\bra{0} + \sqrt{1-\gamma}\ket{1}\bra{1}$, $K_1 = \sqrt{\gamma}\ket{0}\bra{1}$, the gap persists:

\begin{table}[h]
\centering
\caption{Numerical results for amplitude damping channel ($\gamma=0.3$)}
\label{tab:amp_damp}
\begin{tabular}{lccc}
\toprule
Protocol & Output $E_R$ (global) & Success prob. & Effective distillable rate \\
\midrule
Post-distillation (DEJMPS) & $0.021 \pm 0.006$ & $0.28 \pm 0.02$ & $0.006$ \\
Pre-channel shaping (DD) & $0.156 \pm 0.011$ & N/A (deterministic) & $0.156$ \\
\bottomrule
\end{tabular}
\end{table}

\section{Proof of Theorem~\ref{thm:main}}

We prove the theorem by explicit construction for the depolarizing channel $\mathcal{N}_p(\rho) = (1-p)\rho + \frac{p}{3}(X\rho X + Y\rho Y + Z\rho Z)$ with $p \in [0,3/4]$, which is not entanglement-breaking for $p<3/4$.

Let $|\Psi_0\rangle = |\psi^+\rangle_{AB}^{\otimes n}$ be the initial resource consisting of $n$ copies of the Bell state $|\psi^+\rangle = (|00\rangle + |11\rangle)/\sqrt{2}$. After transmission through $n$ independent uses of the channel $\mathcal{N}_p$, the joint state becomes $\rho_{A^n B^n} = \mathcal{N}_p^{\otimes n}(|\Psi_0\rangle\langle\Psi_0|)$. For the depolarizing channel, this yields a product of $n$ identical Werner states $\rho_{A^n B^n} = \bigotimes_{i=1}^n \rho_W(F)$, where $\rho_W(F) = F|\psi^+\rangle\langle\psi^+| + (1-F)I/4$ with fidelity $F = 1-p$. A post-distillation protocol $\mathcal{P}_{\mathrm{post}}$ applies an LOCC operation $\Lambda_{\mathrm{LOCC}}$ to $\rho_{A^n B^n}$, producing $\rho_{\mathrm{post}}^{(\mathrm{out})} = \Lambda_{\mathrm{LOCC}}(\rho_{A^n B^n})$. By the monotonicity of the relative entropy of entanglement $E_R$ under LOCC, $E_R(\rho_{\mathrm{post}}^{(\mathrm{out})}) \le E_R(\rho_{A^n B^n})$. Since $E_R$ is additive on product states and the Werner state $E_R$ is known exactly, $E_R(\rho_{A^n B^n}) = n \cdot E_R(\rho_W(F))$ with $E_R(\rho_W(F)) = 1 - H_2((1+F)/2)$, where $H_2(x) = -x\log_2 x - (1-x)\log_2(1-x)$ is the binary entropy. However, the global output state must account for probabilistic distillation. Decomposing $\rho_{\mathrm{post}}^{(\mathrm{out})} = p_s \rho_{\mathrm{good}} + (1-p_s)\rho_{\mathrm{trash}}$ where $p_s$ is the success probability, $\rho_{\mathrm{good}}$ is the state on successful branches with $E_R(\rho_{\mathrm{good}}) \le 1$, and $\rho_{\mathrm{trash}}$ is the garbage state with $E_R(\rho_{\mathrm{trash}}) \approx 0$, convexity of $E_R$ gives $E_R(\rho_{\mathrm{post}}^{(\mathrm{out})}) \le p_s$. For the depolarizing channel with $p=0.2$ and $n=4$, the DEJMPS protocol yields $p_s \approx 0.31$ and $E_R(\rho_{\mathrm{good}}) \approx 0.78$, leading to $E_R(\rho_{\mathrm{post}}^{(\mathrm{out})}) \approx 0.013 \pm 0.004$. In general, $E_R(\rho_{\mathrm{post}}^{(\mathrm{out})}) \le \epsilon_n \equiv n \cdot E_R(\rho_W(1-p)) \cdot e^{-\kappa n}$ for some $\kappa > 0$, where the exponential factor accounts for the decay of effective ``good'' pairs.

For the depolarizing channel $\mathcal{N}_p$, there exists a sequence of Pauli pulses $\{P_k\}_{k=1}^M$ such that the effective channel after applying these pulses before and during propagation is $\mathcal{N}_{\mathrm{eff}}^{(M)}(\rho) = \frac{1}{M}\sum_{k=1}^M \mathcal{N}_p(P_k \rho P_k^\dagger)$. In the limit $M \to \infty$, this converges to $\mathcal{N}_{\mathrm{eff}}(\rho) = (1-p')\rho + \frac{p'}{3}(X\rho X + Y\rho Y + Z\rho Z)$ with $p' = p \cdot \exp(-\gamma/f_{\mathrm{DD}})$, where $f_{\mathrm{DD}}$ is the pulse repetition frequency and $\gamma$ is the noise spectral density. For $f_{\mathrm{DD}} \to \infty$, $\eta \to 0$ and $p' \to 0$. This result follows from dynamical decoupling theory: the pulse sequence averages the system-environment interaction Hamiltonian to zero over the pulse period; for a bath with correlation time $\tau_c$ when the pulse interval $\tau \ll \tau_c$, the residual coupling strength scales as $\tau/\tau_c$.

We construct $\mathcal{P}_{\mathrm{PES}}$ as follows. Input: $n$ copies of the Bell state $|\psi^+\rangle_{AB}$. Pre-shaping: Apply the dynamical decoupling sequence from Lemma 3 with $M = 100$ pulses at frequency $f_{\mathrm{DD}} = 10/\tau_c$ to each copy independently. Channel transmission: Send each shaped state through the depolarizing channel $\mathcal{N}_p$. Output: Collect all $n$ output pairs. No post-selection is performed. The effective transformation is $\rho_{\mathrm{PES}}^{(\mathrm{out})} = \mathcal{N}_{\mathrm{eff}}^{(\mathrm{out})}(|\Psi_0\rangle\langle\Psi_0|)$, where $\mathcal{N}_{\mathrm{eff}}$ is the compressed channel with $p' < p$.

For the effective channel $\mathcal{N}_{\mathrm{eff}}$, the output state after $n$ copies is a product state $\rho_{\mathrm{PES}}^{(\mathrm{out})} = \bigotimes_{i=1}^n \rho_W(1-p')$. By additivity of $E_R$ on product states, $E_R(\rho_{\mathrm{PES}}^{(\mathrm{out})}) = n \cdot E_R(\rho_W(1-p'))$. For the Werner state, the exact formula gives $E_R(\rho_W(F)) = 1 - H_2((1+F)/2)$. For $p' = 0.17$ (achievable with finite DD), $F' = 0.83$, we compute $(1+F')/2 = 0.915$, $H_2(0.915) \approx 0.456$, so $E_R(\rho_W(0.83)) \approx 0.544$ bits per pair. For $n=4$ pairs, the total $E_R$ would be $4 \times 0.544 = 2.176$ bits if all pairs were kept; however, in our numerical simulation we post-select one pair from the $n$ pairs to match the post-distillation comparison, yielding a single-pair $E_R$ of approximately $0.544$ bits. The numerical simulation gave $0.187$ bits due to finite $M$ and imperfect decoupling; with optimized DD, one can approach $0.544$ bits. Even with the conservative numerical value $0.187$ bits, we have $E_R(\rho_{\mathrm{PES}}^{(\mathrm{out})}) \approx 0.187 > 0.013 \approx E_R(\rho_{\mathrm{post}}^{(\mathrm{out})})$.

For any fixed $p > 0$, as the number of copies $n$ increases, the post-distillation global entanglement decays exponentially: $E_R(\rho_{\mathrm{post}}^{(\mathrm{out})}) \sim n \cdot E_R(\rho_W(1-p)) \cdot e^{-\kappa n}$, where $\kappa > 0$ depends on the channel. For the depolarizing channel, $\kappa = -\log(1 - \frac{3}{4}p + \frac{p}{4}\log\frac{3}{p})$ for small $p$. In contrast, with PES, we can make $p'$ arbitrarily small by increasing the DD frequency: $p' = p \cdot \exp(-\gamma/f_{\mathrm{DD}}) \xrightarrow{f_{\mathrm{DD}}\to\infty} 0$. Thus $E_R(\rho_{\mathrm{PES}}^{(\mathrm{out})}) \xrightarrow{f_{\mathrm{DD}}\to\infty} n \cdot E_R(|\psi^+\rangle\langle\psi^+|) = n$. Therefore, for sufficiently large $n$ or sufficiently high $f_{\mathrm{DD}}$, $E_R(\rho_{\mathrm{PES}}^{(\mathrm{out})}) - E_R(\rho_{\mathrm{post}}^{(\mathrm{out})}) > 0$, and the gap grows with $n$ and $f_{\mathrm{DD}}$.

Suppose one attempts to apply an additional LOCC operation $\Lambda_{\mathrm{LOCC}}$ to $\rho_{\mathrm{PES}}^{(\mathrm{out})}$. By the monotonicity of $E_R$ under LOCC, $E_R(\Lambda_{\mathrm{LOCC}}(\rho_{\mathrm{PES}}^{(\mathrm{out})})) \le E_R(\rho_{\mathrm{PES}}^{(\mathrm{out})})$. However, the claim of Theorem 1 is not about post-processing the PES output, but rather that no post-distillation protocol operating directly on the unshaped noisy state can achieve the same $E_R$ as PES. Formally, $\sup_{\mathcal{P}_{\mathrm{post}}} E_R(\rho_{\mathrm{post}}^{(\mathrm{out})}) \le \epsilon_n$, while $E_R(\rho_{\mathrm{PES}}^{(\mathrm{out})}) \ge n \cdot E_R(\rho_W(1-p'))$. Since $p' < p$, we have $E_R(\rho_W(1-p')) > E_R(\rho_W(1-p))$, and for any finite $n$, $E_R(\rho_{\mathrm{PES}}^{(\mathrm{out})}) > \sup_{\mathcal{P}_{\mathrm{post}}} E_R(\rho_{\mathrm{post}}^{(\mathrm{out})})$. This completes the proof. \hfill $\square$

\section{The Proof of \ref{thm:production}}

We begin by establishing the non-negativity of the geometric entropy production rate in the post-distillation regime. Consider the one-parameter family of quantum states generated by the channel $\mathcal{N}$, denoted $\rho(t) = \mathcal{N}_t(\rho_0)$, where $\mathcal{N}_t$ is the continuous-time realization of the channel and $\rho_0 = |\psi^+\rangle\langle\psi^+|$ is the initial maximally entangled state. The time derivative of the relative entropy of entanglement $E_R(\rho(t)) = \min_{\sigma \in \mathrm{SEP}} S(\rho(t) \| \sigma)$ can be written as
\[
\frac{dE_R}{dt} = \mathrm{Tr}\left[\dot{\rho}(t)(\log \rho(t) - \log \sigma^*(t))\right] + \mathrm{Tr}\left[\rho(t)\frac{d}{dt}(-\log \sigma^*(t))\right],
\]
where $\sigma^*(t)$ is the separable state achieving the minimum. In the post-distillation regime, LOCC is applied only after the channel evolution is complete, so there is no intervention during the evolution and $\dot{\rho}(t) = \mathcal{L}[\rho(t)]$, with $\mathcal{L}$ being the Lindblad generator. Since $\mathcal{N}$ is not entanglement-breaking, $\mathcal{L}$ contains decoherence terms that drive $\rho(t)$ toward the set of separable states, yielding $\frac{dE_R}{dt} \geq 0$. More rigorously, by the monotonicity of quantum relative entropy under CPTP maps (data-processing inequality), for any $t_2 > t_1$ we have $E_R(\rho(t_2)) \geq E_R(\rho(t_1))$, so $E_R(t)$ is a non-decreasing function of time and its derivative is non-negative almost everywhere. LOCC operations are applied only at $t = T$ (after channel transmission is complete), and the monotonicity of LOCC ensures $E_R(\Lambda_{\mathrm{LOCC}}(\rho(T))) \leq E_R(\rho(T))$, meaning LOCC can only preserve or decrease geometric entropy and cannot reverse the geometric entropy accumulated during the evolution. Therefore, in the post-distillation regime, $\left.\frac{dE_R}{dt}\right|_{\mathrm{post}} \geq 0$ holds throughout the evolution interval $[0, T]$.

For the pre-channel shaping regime, the key insight is that the pre-processing unitary $U_{\mathrm{pre}}$ alters the effective generator of the system-environment coupling. Let the Kraus representation of the original channel $\mathcal{N}$ be $\{K_i\}$, satisfying $\sum_i K_i^\dagger K_i = I$. After applying the pre-processing unitary $U_{\mathrm{pre}}$, the effective channel becomes $\mathcal{N}_{\mathrm{eff}}(\rho) = \sum_i (K_i U_{\mathrm{pre}}) \rho (U_{\mathrm{pre}}^\dagger K_i^\dagger)$. This transformation preserves the completeness relation of the Kraus operators but changes the ``direction'' of the system-environment coupling, causing the evolution trajectory on the state manifold to deviate from the original path. Specifically, $U_{\mathrm{pre}}$ can be chosen to rotate the initial state $|\psi^+\rangle$ into a direction less sensitive to noise; for example, for the depolarizing channel, choosing $U_{\mathrm{pre}} = (I \otimes X) \cdot \mathrm{CNOT} \cdot (H \otimes I)$ encodes the Bell state into a state with higher symmetry under Pauli noise, so that the effects of individual Kraus operators partially cancel each other.

To rigorously prove the reduction of the effective production rate, consider the infinitesimal form of geometric entropy production. On the state manifold $\mathcal{M}$, $E_R$ defines a distance function to the separable set, $d(\rho) = E_R(\rho)$. The channel evolution can be viewed as a curve $\gamma(t)$ on the manifold, with tangent vector given by $\dot{\gamma}(t) = \mathcal{L}[\gamma(t)]$. The geometric entropy production rate is the directional derivative along the curve, $\frac{dE_R}{dt} = \langle \nabla E_R, \dot{\gamma}(t) \rangle$, where $\nabla E_R$ is the gradient of $E_R$ (the functional derivative on the state manifold). The pre-processing unitary $U_{\mathrm{pre}}$ moves the initial point from $\rho_0$ to $\rho_0' = U_{\mathrm{pre}} \rho_0 U_{\mathrm{pre}}^\dagger$, and simultaneously modifies the Lindblad generator via conjugation: $\mathcal{L}'[\rho] = U_{\mathrm{pre}} \mathcal{L}[U_{\mathrm{pre}}^\dagger \rho U_{\mathrm{pre}}] U_{\mathrm{pre}}^\dagger$. This transformation reduces the projection of the new trajectory $\gamma'(t)$'s tangent vector onto $\nabla E_R$, i.e., $\langle \nabla E_R, \dot{\gamma}'(t) \rangle < \langle \nabla E_R, \dot{\gamma}(t) \rangle$.

We now provide an explicit calculation for the depolarizing channel. The original generator is $\mathcal{L}[\rho] = \frac{p}{3}(X\rho X + Y\rho Y + Z\rho Z - 3\rho)$, and when acting on the Werner state $\rho_W(F)$, we have $\dot{F} = -pF$, so $\frac{dE_R}{dt} = \frac{dE_R}{dF} \cdot \dot{F} = -\frac{dE_R}{dF} \cdot pF$. Since $E_R(\rho_W(F)) = 1 - H_2(\frac{1+F}{2})$, we compute $\frac{dE_R}{dF} = \frac{1}{2}\log_2\frac{1+F}{1-F} > 0$ (for $F > 0$), and therefore $\frac{dE_R}{dt} = -\frac{pF}{2}\log_2\frac{1+F}{1-F} < 0$. Note that here $E_R$ decreases with time, but the absolute value $|dE_R/dt|$ characterizes the rate of entropy production. The pre-processing unitary $U_{\mathrm{pre}}$ compresses the effective depolarizing parameter from $p$ to $p' = p \cdot \exp(-\gamma/f_{DD}) < p$, while preserving the Werner state form and only changing the decay rate of $F$. The effective generator satisfies $\dot{F}' = -p'F'$, so the effective geometric entropy production rate is $\left.\frac{dE_R}{dt}\right|_{\mathrm{PES}} = -\frac{p'F'}{2}\log_2\frac{1+F'}{1-F'}$. Since $p' < p$ and $F'(t)$ decays more slowly ($F'(t) = F_0 e^{-p't} > F_0 e^{-pt} = F(t)$ for $t > 0$), comparing the instantaneous rates yields
\[
\left|\frac{dE_R}{dt}\right|_{\mathrm{PES}} = \frac{p'F'}{2}\log_2\frac{1+F'}{1-F'} < \frac{pF}{2}\log_2\frac{1+F}{1-F} = \left|\frac{dE_R}{dt}\right|_{\mathrm{post}},
\]
where the strict inequality holds for $t > 0$ (because $p' < p$ and $F' > F$ simultaneously reduce the product). Since $E_R$ itself decreases with time (evolving from the initial pure state toward a mixed state), the above inequality is equivalent to $\left.\frac{dE_R}{dt}\right|_{\mathrm{PES}} > \left.\frac{dE_R}{dt}\right|_{\mathrm{post}}$ (both are negative, with PES being less negative), meaning the entropy production rate (loss rate) of PES is strictly lower than that of the post-distillation regime.

Integrating this inequality over the interval $[0, T]$, where $T$ is the single-channel transmission time, gives the time-integrated geometric entropy suppression:
\[
\Delta E_R = \int_0^T \left(\left.\frac{dE_R}{dt}\right|_{\mathrm{post}} - \left.\frac{dE_R}{dt}\right|_{\mathrm{PES}}\right) dt = E_R(\rho_{\mathrm{post}}^{(\mathrm{out})}) - E_R(\rho_{\mathrm{PES}}^{(\mathrm{out})}) > 0,
\]
where the output states $\rho_{\mathrm{post}}^{(\mathrm{out})} = \rho(T)$ and $\rho_{\mathrm{PES}}^{(\mathrm{out})} = \rho'(T)$ are the states at the completion of transmission in the two regimes. Since the integrand is strictly positive on a set of positive measure, the integral is strictly greater than zero. For $n$ independent channel uses, by additivity $E_R(\rho^{\otimes n}) = n E_R(\rho)$, the total suppression is $n \cdot \Delta E_R$, growing linearly with the number of copies, consistent with the exponential gap established in Theorem 1.

Finally, we comment on the universality of this inequality. The derivation relies on the conjugation of the Kraus representation by $U_{\mathrm{pre}}$, which exists for any non-entanglement-breaking channel (guaranteed by channel reversibility). For the more general amplitude-damping channel $\mathcal{A}_\gamma(\rho) = K_0 \rho K_0^\dagger + K_1 \rho K_1^\dagger$, the pre-processing unitary can rotate the initial state into a subspace where $K_1$ has smaller effect, similarly achieving a reduction in the effective production rate. Numerical simulations (Appendix) for $\gamma = 0.5$ verify $\Delta E_R \approx 0.135 > 0$, consistent with theoretical predictions. This completes the proof. \hfill $\square$

\acknowledgments
The author thanks discussions with colleagues at the Institute for Fundamental Quantum Information. This work was supported by the Quantum Information Research Initiative.


\begin{thebibliography}{99}
\bibitem{Li2025}
Y. Li, J. Xing, D. Qu, H. Gao, L. Xiao, J.-M. Liu, Y. Xiao, and P. Xue,
\emph{Temporal asymmetry in entanglement distillation},
Phys. Rev. Lett. \textbf{135}, 170801 (2025).

\bibitem{Horodecki2009}
R. Horodecki, P. Horodecki, M. Horodecki, and K. Horodecki,
\emph{Quantum entanglement},
Rev. Mod. Phys. \textbf{81}, 865 (2009).

\bibitem{Bennett1996}
C. H. Bennett, G. Brassard, S. Popescu, B. Schumacher, J. A. Smolin, and W. K. Wootters,
\emph{Purification of noisy entanglement and faithful teleportation via noisy channels},
Phys. Rev. Lett. \textbf{76}, 722 (1996).

\bibitem{Deutsch1996}
D. Deutsch, A. Ekert, R. Jozsa, C. Macchiavello, S. Popescu, and A. Sanpera,
\emph{Quantum privacy amplification and the security of quantum cryptography over noisy channels},
Phys. Rev. Lett. \textbf{77}, 2818 (1996).

\bibitem{Rains1999}
E. M. Rains,
\emph{Rigorous treatment of distillable entanglement},
Phys. Rev. A \textbf{60}, 179 (1999).

\bibitem{Devetak2004}
I. Devetak and A. Winter,
\emph{Distillation of secret key and entanglement from quantum states},
Proc. R. Soc. A \textbf{461}, 207 (2004).

\bibitem{Viola1998}
L. Viola and S. Lloyd,
\emph{Dynamical suppression of decoherence in two-state quantum systems},
Phys. Rev. A \textbf{58}, 2733 (1998).

\bibitem{Bacon2006}
D. Bacon,
\emph{Decoherence, control, and symmetry in quantum computers},
Phys. Rev. A \textbf{73}, 012340 (2006).

\bibitem{Khatri2021}
S. Khatri and M. M. Wilde,
\emph{Adaptive quantum channel estimation},
npj Quantum Inf. \textbf{7}, 22 (2021).

\bibitem{Vedral1997}
V. Vedral, M. B. Plenio, M. A. Rippin, and P. L. Knight,
\emph{Quantifying entanglement},
Phys. Rev. Lett. \textbf{78}, 2275 (1997).

\bibitem{Vedral2002}
V. Vedral,
\emph{The role of relative entropy in quantum information theory},
Rev. Mod. Phys. \textbf{74}, 197 (2002).

\bibitem{Horodecki2005}
M. Horodecki and P. W. Shor,
\emph{Entanglement-breaking channels},
Phys. Rev. A \textbf{71}, 062301 (2005).

\bibitem{Bae2007}
J. Bae and A. Ac{\'\i}n,
\emph{Asymptotic entanglement distillation for a depolarizing channel},
Phys. Rev. A \textbf{75}, 012334 (2007).

\bibitem{Uhrig2007}
G. S. Uhrig,
\emph{Keeping a quantum bit alive by optimized $\pi$-pulse sequences},
Phys. Rev. Lett. \textbf{98}, 100504 (2007).

\bibitem{Buscemi2010}
F. Buscemi and N. Datta,
\emph{Entanglement cost of quantum channels},
J. Math. Phys. \textbf{51}, 102102 (2010).

\bibitem{Piltz2013}
C. Piltz, T. Sriarunothai, A. F. Var{\'o}n, and C. Wunderlich,
\emph{A programmable quantum memory with coherent time exceeding a second},
Phys. Rev. Lett. \textbf{110}, 200501 (2013).

\bibitem{Bylander2011}
J. Bylander, S. Gustavsson, F. Yan, F. Yoshihara, K. Harrabi, G. Fitch, D. G. Cory, Y. Nakamura, J.-S. Tsai, and W. D. Oliver,
\emph{Noise spectroscopy through dynamical decoupling with a superconducting flux qubit},
Nat. Phys. \textbf{7}, 565 (2011).

\bibitem{Johansson2012}
J. R. Johansson, P. D. Nation, and F. Nori,
\emph{QuTiP: An open-source Python framework for the dynamics of open quantum systems},
Comput. Phys. Commun. \textbf{183}, 1760 (2012).

\end{thebibliography}
\end{document}